\documentclass[12pt]{article}

\usepackage{latexsym}

\font\tenmsbm=msbm10 scaled 1200
\font\sevenmsbm=msbm9
\newfam\msbmfam
\textfont\msbmfam=\tenmsbm
\scriptfont\msbmfam=\sevenmsbm
\def\msbm{\fam\msbmfam\tenmsbm}

\makeatletter
\@addtoreset{equation}{section}
\makeatother

\newcommand{\eref}[1]{(\ref{#1})}

\def\beq{\begin{equation}}
\def\eeq{\end{equation}}
\def\bea{\begin{eqnarray}}
\def\eea{\end{eqnarray}}
\def\bet{\begin{tabular}}
\def\eet{\end{tabular}}
\def\ol{\overline}

\def\a{\alpha}
\def\b{\beta}
\def\l{\Lambda}
\def\f{\phi}

\def\g{\gamma}

\def\e{\varepsilon}

\def\G{\Gamma}
\def\D{{\partial}}

\def\bl{\bar{\Lambda}}

\def\bpsi{\bar{\psi}}
\def\bg{\bar{g}}

\def\be{\bar{\epsilon}}
\def\bff{\bar{f}}
\def\bu{\ol{U}}
\def\bv{\ol{V}}
\def\bA{\bar{A}}
\def\bF{\bar{F}}

\def\bR{\bar{R}}
\def\bS{\bar{S}}
\def\bC{\bar{C}}

\def\tC{\tilde{C}}

\def\hA{\hat{A}}
\def\hS{\hat{S}}
\def\hR{\hat{R}}
\def\hF{\hat{F}}
\def\hC{\hat{C}}
\def\hg{\hat{g}}
\def\hf{\hat{f}}

\catcode`@=11
\def\lsim{\mathchoice
  {\mathrel{\lower.8ex\hbox{$\displaystyle\buildrel<\over\sim$}}}
  {\mathrel{\lower.8ex\hbox{$\textstyle\buildrel<\over\sim$}}}
  {\mathrel{\lower.8ex\hbox{$\scriptstyle\buildrel<\over\sim$}}}
  {\mathrel{\lower.8ex\hbox{$\scriptscriptstyle\buildrel<\over\sim$}}} }
\def\gsim{\mathchoice
  {\mathrel{\lower.8ex\hbox{$\displaystyle\buildrel>\over\sim$}}}
  {\mathrel{\lower.8ex\hbox{$\textstyle\buildrel>\over\sim$}}}
  {\mathrel{\lower.8ex\hbox{$\scriptstyle\buildrel>\over\sim$}}}
  {\mathrel{\lower.8ex\hbox{$\scriptscriptstyle\buildrel>\over\sim$}}} }
\def\croce{\displaystyle / \kern-0.2truecm\hbox{$\backslash$}}
\def\lqua{\lower4pt\hbox{\kern5pt\hbox{$\sim$}}\raise1pt
\hbox{\kern-8pt\hbox{$<$}}~}
\def\gqua{\lower4pt\hbox{\kern5pt\hbox{$\sim$}}\raise1pt
\hbox{\kern-8pt\hbox{$>$}}~}
\def\mma{\lower1pt\hbox{\kern5pt\hbox{$\scriptstyle <$}}\raise2pt
\hbox{\kern-7pt\hbox{$\scriptstyle >$}}~}
\def\mmb{\lower1pt\hbox{\kern5pt\hbox{$\scriptstyle >$}}\raise2pt
\hbox{\kern-7pt\hbox{$\scriptstyle <$}}~}
\def\mmc{\lower4pt\hbox{\kern5pt\hbox{$<$}}\raise1pt
\hbox{\kern-8pt\hbox{$>$}}~}
\def\mmd{\lower4pt\hbox{\kern5pt\hbox{$>$}}\raise1pt
\hbox{\kern-8pt\hbox{$<$}}~}
\def\lsu{\raise4pt\hbox{\kern5pt\hbox{$\sim$}}\lower1pt
\hbox{\kern-8pt\hbox{$<$}}~}
\def\gsu{\raise4pt\hbox{\kern5pt\hbox{$\sim$}}\lower1pt
\hbox{\kern-8pt\hbox{$>$}}~}
\def\croce{\displaystyle / \kern-0.2truecm\hbox{$\backslash$}}
\def\ali{\hbox{A \kern-.9em\raise1.7ex\hbox{$\scriptstyle \circ$}}}
\def\2frecce{\hbox{\lower 0.3ex\hbox{$\leftarrow$} 
\hbox{\kern-1.3em\raise 0.3ex\hbox{$\rightarrow$}}}}
\def\quad@rato#1#2{{\vcenter{\vbox{
        \hrule height#2pt
        \hbox{\vrule width#2pt height#1pt \kern#1pt \vrule width#2pt}
        \hrule height#2pt} }}}
\def\quadratello{\mathchoice
\quad@rato5{.5}\quad@rato5{.5}\quad@rato{3.5}{.35}\quad@rato{2.5}{.25} }

\begin{document}

\begin{titlepage}

\begin{flushright}
Preprint DFPD 98/TH/29\\
DFTT 29/98 \\
hep-th/9806140\\
June 1998\\
\end{flushright}

\vspace{2truecm}

\begin{center}

{ \Large \bf $D=10$, $N = IIB$  Supergravity: Lorentz--invariant 
actions and duality}

\vspace{1cm}

{ Gianguido Dall'Agata$^{\dag}$}\footnote{dallagat@to.infn.it}, { 
Kurt 
Lechner$^{\ddag}$}\footnote{kurt.lechner@pd.infn.it}  and
{Mario Tonin$^\ddag$}\footnote{mario.tonin@pd.infn.it} 
\vspace{1cm}

{$\dag$ \it Dipartimento di Fisica Teorica, Universit\`a degli 
studi di Torino, \\
 via P. Giuria 1,I-10125 Torino}

\medskip

{ $\ddag$ \it Dipartimento di Fisica, Universit\`a degli Studi di 
Padova,

\smallskip

and

\smallskip

Istituto Nazionale di Fisica Nucleare, Sezione di Padova, 

Via F. Marzolo, 8, 35131 Padova, Italia
}

\vspace{1cm}

\begin{abstract}

We present a manifestly Lorentz invariant and supersymmetric 
component 
field action for $D = 10$, type $IIB$ supergravity, using a newly 
developed method for the construction of actions with chiral bosons, 
which implies only a single scalar non propagating auxiliary field.
With the same method we construct also an action in which the complex 
two--form gauge potential and its Hodge--dual, a complex six--form 
gauge potential, appear in a symmetric way in compatibility with 
supersymmetry and Lorentz invariance.
The duals of the two physical scalars of the theory turn out to be 
described by a $SL(2, {\msbm R})$ triplet of eight--forms whose 
curvatures are constrained by a single linear relation.
 We present also a supersymmetric action in which the basic 
fields and their duals, six--form and eight--form potentials, appear 
in a symmetric way.
All these actions are manifestly invariant under the global 
$SL(2,{\msbm R})$--duality group of $D = 10$, $IIB$ supergravity and 
are 
equivalent to each other in that their dynamics corresponds to the 
well known equations of motion of $ D=10$, $IIB$ supergravity.

\end{abstract}

\end{center}
\end{titlepage}

\newpage

\baselineskip 6 mm


\section{Introduction and Summary}

The obstacle which prevented for a long time a Lagrangian formulation 
of $D=10$, $N=IIB$ supergravity is the appearance of a chiral boson 
in the spectrum of the theory, i.e. a four--form gauge potential with 
a self--dual field strength.
A possible way to overcome this obstacle was presented in 
\cite{armeni}, by extending Siegel's action for two--dimensional
chiral bosons \cite{Siegel} to higher dimensions. 
In this approach one gets, through lagrangian multipliers, not the
self--duality condition as equation of motion for the chiral 
bosons, but rather its square.
However, in dimensions greater than two, the elimination of the 
lagrangian multipliers seems problematic \cite{Siegel}
and, moreover, at the quantum level, for example in the derivation 
of the Lorentz anomaly, it seems to present untreatable
technical and conceptual problems. On the other hand, a group manifold 
action for $D=10,IIB$ supergravity has been obtained in \cite{Cast}.
In general, when a group manifold action is restricted to 
ordinary space--time one gets a consistent supersymmetric 
action for the component fields; but if a self--dual (or anti self--dual) 
tensor is present, the  restricted action loses supersymmetry \cite{fre}.

A new method for writing manifestly Lorentz--invariant and 
supersymmetric actions for chiral bosons ($p$--forms)
in $D = 2 \hbox{ mod } 4$ 
has been presented in \cite{PST}: it uses a single non 
propagating auxiliary scalar field and involves two new bosonic 
symmetries; 
one of them allows to eliminate the auxiliary field and the other 
kills half of the degrees of freedom of the $p$--form, reducing it to 
a chiral boson.
This method turned out to be compatible with all relevant symmetries, 
including supersymmetry and $\kappa$--symmetry \cite{DLT,M5} and 
admits also a canonical coupling to gravity, being manifestly Lorentz 
invariant. As all lagrangian formulations of theories with chiral
bosons, the method is expected to be
insufficient for what concerns the quantization of these actions on 
manifolds with non--trivial topology
\cite{Witten}, but it can be successfully applied, even at the quantum 
level, on trivial manifolds. 
As an example of the efficiency of this method also at the quantum
level, we mention that the effective 
action for chiral bosons in two dimensions, coupled to a 
background metric, can easily be computed in a covariant way 
\cite{KL}, and that it gives the expected result, namely
the effective action of a two--dimensional complex Weyl--fermion.
This implies, in turn,  that also the Lorentz-- and 
Weyl--anomalies due to a $D = 
2$ chiral boson, as derived with this new method, coincide with the ones 
predicted by the index theorem. 
Work regarding the Lorentz--anomaly in higher dimensions is in 
progress.

In this paper we present a manifestly 
Lorentz--invariant and supersymmetric action for $D=10$, $N = IIB$ 
supergravity, based on this method\footnote{The covariant action for 
the bosonic sector of type $IIB$ supergravity has already been 
presented in \cite{IIB}.}. 
Apart from the above mentioned  new features, the basic ingredients 
are 
the equations of motion and SUSY--transformations of the basic 
fields, which are well known \cite{HW, SW, SC}, and can be most 
conveniently derived in a superspace approach \cite{HW,Cast}.
In addition to the metric, the bosonic fields in this theory are two 
complex scalars ($0$--forms), which parametrize the coset 
$\displaystyle \frac{SL(2, {\msbm R})}{U(1)}$, a complex two--form 
gauge potential 
and 
the real chiral four--form gauge potential.

Since the discovery of $D$--branes and their coupling to RR gauge 
potentials, the Hodge duals to the zero-- and two--forms (the 
four--form is self--dual) i.e. the eight-- and six--forms acquired a 
deeper physical meaning.
It is therefore of some interest to look at a Lagrangian 
formulation with manifest duality i.e. in which the zero and eight--forms
 and the two and 
six--forms appear in a symmetric way. The method presented in \cite{PST}
appears particularly suitable to cope also with this problem.
Indeed, a variant \cite{Maxwell,D11} of this method
allowed in the past to  construct a manifestly duality invariant Lagrangian 
for Maxwell's equations in four dimensions \cite{Maxwell} as well as for 
$N=1,D=11$ supergravity \cite{D11}.

In this paper 
we shall also construct an action with manifest duality between the 
gauge potentials and their Hodge duals. Upon gauge fixing the new bosonic
symmetries, mentioned above, one can remove the six-- and 
eight--forms and recover the (standard) formulation 
with only zero-- and two--forms.
On the other hand, a Lagrangian formulation in which only the six 
and eight--forms appear, instead of the zero and two--forms, is  not 
accessible for intrinsic reasons i.e. the presence of Chern--Simons
forms in the definition of the curvatures.

The case of the eight--forms requires a second variant of 
the method.
As we will see, manifest invariance under the global $S$--duality 
group $SL(2,{\msbm R})$ of the theory requires the introduction of 
{\it three} 
real eight--form potentials, with three nine--form curvatures, which 
belong to the adjoint representation of $SL(2,{\msbm R})$.
Two $SL(2,{\msbm R})$--invariant combinations of the three nine--form 
curvatures are related by Hodge--duality to the two real (or one 
complex) one--form curvatures of the scalars.
The third one is determined by an 
$SL(2,{\msbm R})$--invariant linear constraint between the three 
curvature nine--forms.
While the first variant of our method allows to treat {\it 
Hodge--duality relations} between forms at a Lagrangian level, the 
second variant allows to deal, still at a Lagrangian level, with {\it 
linear 
relations} between curvatures.
This is then precisely what is needed to describe the dynamics of the 
eight--forms through an action principle.

The general validity of the method is underlined also by the fact 
that 
all these lagrangians are supersymmetric.
To achieve supersymmetry one has to modify the SUSY transformation 
laws of the fermions in a very simple and canonical way, the 
modifications being proportional to the equations of motion, derived
e.g. in a superspace approach.
The on--shell SUSY algebra can then be seen to close on the two new 
bosonic symmetries mentioned at the beginning.

In section two we  present the superspace language and results 
 for $D = 10$, $N = IIB$  supergravity, following mainly \cite{HW}.
There, we construct also the dual supercurvatures and potentials in 
a $SL(2,{\msbm R})$ covariant way. 
These results are used at the {\it component level}, in section four, 
to write a manifestly Lorentz--invariant action for the theory, using 
only the scalars, two and four--forms.
In section three we give a concise account of the new method itself 
(for more details see \cite{PST, DLT}).
In section five we write an action in which the two and six--form 
potentials appear in a symmetric way and prove its invariance 
under supersymmetry.
Section six is devoted to the construction of an action in which all 
gauge potentials appear in a symmetric way, paying special attention 
to the new features exhibited by the eight--forms.
Section seven collects some concluding remarks and observations.

\section{Superspace results}

The superspace conventions and  results of this paper follow 
mainly \cite{HW}.
The $D = 10$, $IIB$ superspace is parametrized by the 
supercoordinates $Z^M = ( x^m, \theta^\mu, \bar{\theta}^\mu)$ where 
the $\theta^\mu$ are sixteen complex anticommuting coordinates.
Here and in what follows the "bar" indicates simply complex 
conjugation and in case transposition.
The cotangent superspace basis is indicated by $e^A = d Z^M 
{e_M}^A(Z) = (e^a, e^{\a}, \bar{e^{\a}}) \equiv (e^a, \psi^{\a}, 
\ol{\psi^{\a}})$, where $a = 0,1, \ldots, 9$ and $\a = 1, \ldots, 
16$, and $\psi^{\a} = d Z^M {\psi_M}^{\a}$ indicates the complex 
gravitino one--superform.
All superforms can be decomposed along this basis. 
The Lorentz superconnection one--form is given by ${\omega_a}^b = d 
Z^M {\omega_{M\,a}}^b$ with curvature ${R_a}^b = d {\omega_a}^b + 
{\omega_a}^c {\omega_c}^b$.

 The two physical real scalars of the theory parametrize the coset 
$\displaystyle \frac{SU(1,1)}{U(1)}$, where $SU(1,1) 
\approx SL(2, {\msbm R})$  is the global $S$--duality symmetry group 
of 
$D=10$, 
$IIB$ supergravity, and the $U(1)$ is realized locally.
The coset is described by two complex scalars $(U, -\bv) \equiv 
A_{0}$ 
which are constrained by 
$|U|^{2} - |V|^2 = 1$ such that the matrix 
\beq
\label{0}
 W \equiv \left( \bet{cc} $U$ & 
$V$ \\ $\bv$ & $\bu$ \eet \right) 
\eeq
belongs to $SU(1,1)$ and the 
fields $ A_{0} \equiv (U, -\bv)$ form an 
$SU(1,1)$ doublet.
The Maurer--Cartan form $W^{-1} dW$ decomposes then as 
\beq 
\label{2aa}
W^{-1} d W 
= \left( \bet{cc} $2iQ$ & $R_1$ \\ 
$\bR_{1}$ & $-2iQ$ \eet \right) 
\eeq
where $R_{1}$ and $Q$ are 
$SU(1,1)$ invariant one--forms:
\bea
R_1 &=& \bu dV - V d \bu, \\
Q &=& \frac{1}{2i} ( \bu dU - V d \bv).
\eea
Since the $U(1)$ weights of $(U,V)$ are $(-2,2)$, i.e. $A_0$ has 
weight $-2$, $Q$ is a $U(1)$--connection 
and $R_1$, which has to be considered as the curvature of 
the scalars, has $U(1)$ weight 4.
We can then introduce a $U(1)$ and $SO(1,9)$ covariant derivative 
which acts on a $p$--form with $U(1)$ weight $q$ as
\beq
D = d + \omega + q i Q.
\eeq
For a list of the $U(1)$ weights and $SU(1,1)$ representations of the 
fields see the table at the end of the section.

For a $p$--form with purely bosonic components $$ \phi_{p} = 
\frac{1}{p!} e^{a_1} \ldots e^{a_{p}} \phi_{a_{p} \ldots a_1}, $$ we 
introduce its Hodge--dual, a $(10-p)$--form, as
\beq
*(\phi_{p}) \equiv \frac{1}{(10 - p)!} e^{a_1} \ldots e^{a_{10-p}} 
(*\phi)_{a_{10-p} \ldots a_1},
\eeq
where 
$$
(*\phi)_{a_1 \ldots a_{10-p}} \equiv \frac{1}{p!} \epsilon_{a_1 
\ldots 
a_{10-p}}{}^{b_1\ldots b_p} \phi_{b_1 \ldots b_p}. $$
In particular, on a $p$ --form we have $$ (*)^{2} = (-1)^{p+1}. $$

The other bosonic degrees of freedom are carried by the following 
superforms.
We introduce a complex two--form $A_2$, where $(A_2, \bA_2)$ 
constitute an $SU(1,1)$ doublet, and its dual which is a complex 
six--form $A_6$, where $(iA_6, \ol{iA_{6}})$ constitutes also a 
$SU(1,1)$ doublet.
The real four--form in the theory, the "chiral boson", is denoted by 
$A_4$.
As anticipated in the introduction the duals of the scalars are 
parametrized by three real eight--forms which are described by a 
complex eight--form $A_8$ and a purely imaginary one $\hA_8$.
The three forms $\left(A_8, \bA_8, \hA_8\right)$ form an 
$SU(1,1)$ triplet, i.e.  
they belong to the adjoint representation of $SU(1,1)$.  All these 
forms are $U(1)$ singlets.

The curvatures associated to these forms maintain their $SU(1,1)$ and 
$U(1)$ representations and are given by
\bea
\label{1a}
S_3 = d A_2, && dS_3 = 0, \\
\label{1b}
S_5 = d A_4 + i (A_2 d \bA_2 - \bA_2 d A_2), && d S_5 = 2i \bS_{3} 
S_3,\\
\label{1c}
S_7 = \displaystyle d A_6 - \frac{i}{3} (S_5 + 2d A_4) A_2, && d S_7 = 
i S_3 S_5,\\
\label{1d}
S_9 = \displaystyle d A_8 + \left[ \bS_7 - \frac{i}{4} \bA_2 ( S_5 + 
dA_4)\right] 
\bA_2, && dS_{9} = \bS_7 \bS_3,\\
\label{1e}
\hS_9 = \displaystyle  d \hA_8 + \frac{1}{2} \left[ S_7 + 
\frac{i}{4} A_2 ( S_5 
+ dA_4)\right] \bA_2 - c.c.,  && d \hS_9 = \frac{1}{2} ( S_7 \bS_3 - 
\bS_7 S_3). 
\eea
Again $\left(S_9, \bS_9, \hS_9\right)$ form an $SU(1,1)$ triplet.

The superspace parametrizations of these curvatures are more 
conveniently given in terms of the $SU(1,1)$ invariant combinations 
which one can form using the scalars $U$ and $V$. 
Including also the curvatures for the scalars these invariant 
curvatures are given by
\bea
\label{2a}
R_1 &=& \bu dV - V d\bu, \\
\label{2b}
R_3 &=& \bu S_3 + V \bS_3, \\
\label{2c}
R_5 &=& S_5, \\
\label{2d}
R_7 &=& \bu S_7 - V \bS_7, \\
\label{2e}
R_9 &=& \bu^2 \bS_9 - V^2 S_9 + 2 \bu V \hS_9, \\
\label{2f}
\hR_9 &=& UV S_9 - \bu \bv \bS_9 - \left( |U|^2 + |V|^2\right) 
\hS_9.
\eea

$R_1$ and $R_9$ carry $U(1)$ charge 4, $R_3$ and $R_7$ carry charge 2 
and $R_5$ and $\hR_9$ carry charge 0 and are respectively real 
and 
purely imaginary, while all other $R_n$ are complex.
The associated Bianchi identities are
\bea
\label{3a}
D R_1 &=& 0 \\
\label{3b}
D R_3 &=& \bR_3 R_1, \\
D R_5 &=& 2i \bR_3 R_3, \\
\label{3d}
D R_7 &=& -\bR_7 R_1 + i R_3 R_5, \\
\label{3e}
D R_9 &=& 2 R_1 \hR_9 + R_7 R_3, \\
\label{3f}
D \hR_9 &=& \bR_1 R_9 - R_1 \bR_9 + \frac{1}{2} \left( \bR_7 R_3 
- R_7 
\bR_3\right).
\eea
For the $U(1)$ connection we have
\beq
d Q = \frac{i}{2} \bR_1 R_1;
\eeq
it is also useful to notice that
\bea
DU &=& V \bR_1, \nonumber \\
\label{3i}
DV &=& U R_1.
\eea

Defining the torsion as usual by
\beq
T^A = D e^A,
\eeq
it satisfies the Bianchi identities
\bea
D T^{\a} &=& \psi^{\g} {R_{\g}}^{\a} + \frac{1}{2} \bR_1 R_1 
\psi^{\a}, \\
D T^a &=& e^b {R_{b}}^a.
\eea

The superspace parametrizations of the curvatures in 
\eref{2a}--\eref{2f} can 
now be written, for $n = 1,3,5,7,9,\hat{9}$, as
\beq
\label{5}
R_n = F_n - C_n,
\eeq
where $F_n$ indicates the purely bosonic part
\beq
F_n = \frac{1}{n!} e^{a_1} \ldots e^{a_n} F_{a_n \ldots a_1},
\eeq
and the $n$--forms $C_n$ involve the gravitino one--form $\psi^{\a}$ 
and the complex spinor $\l_{\a}$, which completes the fermionic 
degrees of freedom of $D = 10$, $IIB$ supergravity (contraction of 
spinorial indices is understood):
\bea
\label{7}
C_1 &=& 2 \psi \l, \\
C_3 &=& \frac{1}{2} e^a e^b \left( \bpsi \G_{ab} \l\right) + 
\frac{i}{2} e^a 
(\psi \G_{a} \psi), \\\
\label{7b}
C_5 &=& - \frac{1}{3!} e^a e^b e^c \left( \bpsi \G_{abc}\psi\right) + 
\frac{1}{5!} e^{a_1} \ldots e^{a_5} \left(\bl \G_{a_1 \ldots a_5} 
\l\right), \\
C_7 &=& \frac{1}{6!} e^{a_1} \ldots e^{a_6} \; \left(  \bar{\psi}
\Gamma_{a_1 \ldots a_6} \Lambda \right) - \frac{i}{2} \frac{1}{5!} \;
e^{a_1} \ldots e^{a_5} \left(\psi \Gamma_{a_1 \ldots a_5} \psi \right), \\
C_9 &=& \frac{2}{8!} e^{a_1} \ldots e^{a_8} (\psi \G_{a_1 \ldots 
a_8 } \l),\\
\label{7d}
\hC_9 &=& -\frac{i}{2} \left[\frac{e^{a_1} \ldots e^{a_7}}{7!} 
\left( \bar{\psi} \G_{a_1 
\ldots a_7} \psi\right) + \frac{6}{9!} e^{a_1} \ldots e^{a_9} \left( 
\bl 
\G_{a_1 \ldots a_9} \l\right) \right]. 
\eea

Actually, $C_5$ and $\hC_9$ contain also a contribution with 
only bosonic vielbeins. These amount, however, only to a redefinition of 
$F_5$ and $\hF_9$. 
These redefinitions are convenient for what follows, see eqs. 
\eref{13}--\eref{16} below.
It is also convenient to decompose the forms $C_n$ as
\beq
C_n = C_n^{\Lambda} + C_n^{\psi}
\eeq
where $C_n^{\Lambda}$ indicates the parts which depend on $\l_{\a}$ 
and $ C_n^{\psi}$ the parts which are independent of $\l_{\a}$, in 
particular $C_1^{\psi} = 0 = C_9^{\psi}$.

The parametrizations of the torsions and of $D\l_{\a}$ become
\bea
\label{9}
D e^a &=& i \bpsi \G^a \psi, \\
\label{10}
D \psi &=& \frac{1}{2} e^a e^b T_{ba} - (\bpsi \l) \bpsi - 
\frac{1}{2} ( \bpsi \G^a \bpsi) \G^a \l + \nonumber \\
&+& i e^a\left[ -\frac{21}{2} X_{a} + \frac{3}{2} X^b \G_{ab} + 
\frac{5}{4} X_{abc }\G^{bc} - \frac{1}{4} X^{bcd} \G_{abcd} + \right. 
\nonumber \\
&+& \left. \frac{1}{192} ( F^{+}_{abcde}- 4 X_{abcde}) 
\G^{bcde}\right]\psi + \frac{3}{16} e^a \left( - F_{abc}\G^{bc} + 
\frac{1}{9} F^{bcd} \G_{abcd}\right) \bpsi, \\
\label{11}
D \l &=& e^b D_b \l + \frac{i}{2} F^a \G_a \bpsi + \frac{i}{24} 
F^{abc} \G_{abc} \psi.
\eea
Here $T_{ab}{}^{\a}$ parametrizes the part of $D \psi^{\a}$ with only 
bosonic vielbeins and
\beq
X^{(n)} \equiv \frac{1}{16} \bl \G^{(n)} \l.
\eeq
$F^+_{a_1 \ldots a_5}$ indicates the self--dual part of $F_{a_1 
\ldots 
a_5}$:
\beq
F^{\pm}_{a_1 \ldots a_5} = \frac{1}{2} ( F_{a_1 \ldots a_5} \pm (*F)_{a_1 
\ldots a_5}).
\eeq
For the parametrization of $R_a{}^b$, see \cite{HW}.

All these parametrizations and the form of the Bianchi identities are 
dictated by the consistency of the Bianchi identities for the torsion 
and for the curvatures $R_n$ themselves.

Since the closure of the SUSY--algebra sets the theory on shell one 
gets also the following (self)--duality relations between the 
curvatures, which become extremely simple when expressed through the 
$F_n$ defined in \eref{5}-\eref{7d}:
\bea
\label{13}
* F_5 &=& F_5, \\
\label{14}
* F_7 &=& F_3, \\ 
\label{15}
* F_9 &=& F_1, \\
\label{16}
* \hF_9 &=& 0.
\eea
The first relation is the equation of motion for the four--form 
$A_4$. Equation \eref{14}, which relates the curvature of $A_6$ to 
the curvature of $A_2$, promotes the Bianchi identities \eref{3b} 
and \eref{3d} to equations of motion for $A_6$ and $A_2$ 
,respectively.

Equation \eref{16} constitutes the linear constraint between $d\hA_8$ 
and $dA_8$ mentioned in the introduction, and allows to 
express --- through \eref{2f} --- the curvature $\hS_9$, and 
hence $\hA_8$, as a function of $S_9$.
Substituting this expression for $\hS_9$ in \eref{2e} one can 
compute $R_9$ and $F_9$ as a function of $dA_8$.
At this point the duality relation \eref{15} promotes \eref{3a} and 
\eref{3e} to equations of motion for $A_8$ and $A_{0}= (U,-\bv)$ 
respectively.
The complex eight--form $A_8$ is thus dual to the two real scalars 
contained in $U$ and $V$.
These fields are,in fact, constrained by $|U|^2 - |V|^2 = 1$ and are subjected 
to the local $U(1)$ invariance. 
Once this invariance is fixed only two real physical scalars survive.
It is clear that the elimination of $\hA_8$ breaks manifest 
$SU(1,1)$ invariance and that a manifestly $SU(1,1)$ invariant action 
principle for the dual scalars has to be based on three eight--forms, 
i.e. \eref{1d}--\eref{1e}, \eref{2e}--\eref{2f} and 
\eref{15}--\eref{16}.

The occurrence of three eight--forms can also be understood from the 
following point of view.
Since the theory possesses a global $SU(1,1)$ invariance there must 
exist three conserved currents which belong to the adjoint 
representation of $SU(1,1)$.
The Hodge duals of these currents, which have to be closed and hence 
locally exact, are just given by $dA_8$ and $d\hA_8$ and their 
explicit expressions can be derived from \eref{1d} and \eref{1e} 
using \eref{15} and \eref{16}.

An expression for the dual curvatures and their Bianchi identities was 
given also in \cite{ceder}, in a non--manifestly  $SU(1,1)$ and $U(1)$ 
covariant formulation. In this formulation it is sufficient to
introduce only {\it two} eight--form potentials because the $U(1)$ invariance 
has been gauge fixed. 

The equations of motion for the gravitino, for $\l_{\a}$ and for the 
metric can be found in \cite{HW}; their explicit expressions are not 
needed here since the action is completely determined by SUSY 
invariance, by the knowledge of the Bianchi identities and by the 
superspace parametrizations given above.

The $SU(1,1)$ and $U(1)$ representations of the basic fields and 
their 
charges are:
\begin{center}
\begin{tabular}{|c|c|c|c|c|c|c|c|c|c|}\hline
$\phantom{\stackrel{A}{B}}$ & $e^a$ & $\psi^{\a}$ & $\l_{\a}$ & $A_0$ 
& $\left(A_{2,6}, 
\bA_{2,6}\right)$ & 
 $\left(A_8, \bA_8, \hA_8\right)$ &$R_{5}, \hR_9$ & $R_3$, 
$R_7$ & 
 $R_1$, $R_9$ \\\hline
$\phantom{\stackrel{A}{B}} n[U(1)] \phantom{p_j}$ &  0 & 1 & 3 & -2 & 
0 & 0 & 0 & 2 & 4 \\\hline
$\phantom{\stackrel{A}{B}} SU(1,1) \phantom{j_j}$ & 1 & 1 & 1 & 2 & 2 
& 3 & 1 & 1 & 1 \\\hline
\end{tabular}
\end{center}

\section{The covariant method}

From now on we will work in {\it ordinary} space--time but still 
continue to use the language of forms to avoid the explicit 
appearance of Lorentz indices. 
In particular, our actions will be written as integrals over 
ten--forms.
In the next section we will perform the reduction of the superspace 
results of the preceding section to ordinary bosonic space--time.

In this section we will present the basic ingredients which allow to 
write covariant actions for equations \eref{13}-\eref{16} 
concentrating in 
particular on the self--duality equation of motion \eref{13}.

This equation is of the type
\beq
\label{17}
F^-_5 \equiv \frac{1}{2} ( F_5 - * F_5) = 0,
\eeq
where 
\beq
F_5 = d A_4 + \tC_5
\eeq
and $\tC_5$ is independent of $A_4$\footnote{In the case of $D = 10$, 
$IIB$ supergravity we have $\tC_5 = C_5 + i ( A_2 d \bA_2 - \bA_2 d 
A_2)$, now in ordinary space--time.}.

The covariant method \cite{PST} requires the introduction of a scalar 
auxiliary field $a(x)$ and the related vector
\beq
v_{a}(x) = \frac{{e_a}^m \D_m a}{\sqrt{-g^{mn}\D_m a\D_n a}}\equiv 
{e_{a}}^m v_m,
\eeq
satisfying $v^a v_a = -1$.
We introduce also the one--form 
\beq
v = e^a v_a = {da\over 
\sqrt{-g^{mn}\D_m a\D_n a}},
\eeq
and indicate with $i_v$ 
the interior product of a $p$--form with the vector field $v^m \D_m$.

Defining 
\beq
f_4 \equiv i_v (F_5 - * F_5),
\eeq
the action which reproduces \eref{17} can be written as
\bea
S_0[A_4, a] &=& \frac{1}{2} \int \left[ \frac{1}{2} \left( F_5 * F_5 
+ f_4 * 
f_4 \right) +  \tC_5 d A_4 \right]  \nonumber \\
&=& \frac{1}{2} \int \left[ \frac{1}{2} \left( F_5 * F_5 + f_4 * 
f_4  \right) + F_5 d A_4 \right].
\eea
The form of this action is selected, and fixed, by the following 
symmetries:
\bea
I) & & \delta A_4 = \Lambda_3 \, da,  \qquad \qquad \delta a = 0,\\
II) & & \delta A_4 = - \frac{\f}{\sqrt{- (\D a)^2}} f_4 \quad \delta 
a = \f, 
\eea
where $\f$ and $\Lambda_3$ are transformation parameters, 
respectively a scalar and a three--form.
The action $S_0$ is, actually, invariant also under {\it finite} 
transformations of the type $I)$ i.e. under $A_4 \to A_4 + 
\Lambda_3Ê\, da $. This fact becomes relevant in what follows.

The equation of motion for $a$ and $A_4$  are respectively given by
\bea
\label{21}
d \left( \frac{1}{\sqrt{- ( \D a)^2}} v f_4 f_4 \right) &=& 0, \\
\label{22}
d (v f_4 ) &=& 0,
\eea

The symmetry $II)$ promotes the auxiliary field $a$ to a "pure gauge" 
field and allows to gauge--fix\footnote{Typical non--covariant gauges 
are $a_0(x) = n_m x^m$ 
where $n_m$ is a constant vector.} it to an arbitrary function $a(x) 
= 
a_0(x)$ provided that $g^{mn} \D_m a_0 \, \D_n a_0 \neq 
0$. Correspondingly, the equation of 
motion \eref{21} can easily be seen to be a consequence of \eref{22}.

The general solution of \eref{22} is $v f_4 = d \tilde{\Lambda}_3 \, 
da $. 
Since under a finite transformation $I)$ we have 
$$ 
v f_4 \to v f_4 +  d \Lambda_3 \, da,
$$
choosing $\Lambda_3 = \tilde{\Lambda}_{3}$ we get
\beq
\label{23}
f_4 = 0.
\eeq

Due to the identity decomposition on a $p$--form
\beq
\label{24}
{\msbm I} = (-1)^p v i_v - * v i_{v} *,
\eeq
one gets the identity
\beq
\label{25}
F_5 - * F_5 = - v f_4 + * (v f_4)
\eeq
and hence \eref{23} is equivalent to the self--duality equation of 
motion for $A_4$ \eref{17}.

This concludes the proof that the action $S_0$ describes indeed 
interacting ($\tC_5 \neq 0$) chiral bosons in ten dimensions.
If the fields composing $\tC_5$ are themselves dynamical, one has to 
complete the action $S_0$ by adding terms which involve the kinetic 
and interaction terms for those fields, but not $A_4$ itself because 
otherwise the symmetries $I)$ and $II)$ are destroyed.

Another five--form, which will acquire an important role  
 in establishing supersymmetry invariance, is given by 
\beq
\label{26}
K_5 \equiv F_5 + v f_4.
\eeq

This five--form is uniquely determined by the following properties:
it is self--dual,
$$
K_5 = * K_5,
$$
as follows  from \eref{25}, and it reduces to $F^+_5$ if the self--duality
constraint for $A_4$ \eref{17} holds. 
$K_5$ constitutes therefore a kind of off--shell generalization of 
$F^+_5$.

\section{The complete action for $D = 10$, $IIB$ supergravity}

In this section we write a covariant and supersymmetric component 
level action for $IIB$ supergravity in its canonical formulation, 
i.e. when the bosonic degrees of freedom are described by $A_0$, 
$A_2$ and $A_4$, incorporating the dynamics of $A_4$ according to the 
method presented in the preceding section.

The component results are obtained from the superspace results of 
section two in a standard fashion setting $ \theta = 0 = d \theta$.
Whenever we use the same symbols as in section two we mean those 
objects evaluated at  $ \theta = 0 = d \theta$.
In particular the differential $d$ becomes the ordinary differential.
Every form can now be decomposed along the vielbeins $e^a = dx^m 
{e_m}^a$ and the gravitino reduces to $\psi^{\a} = dx^m 
{\psi_m}^{\a} \equiv e^a {\psi_a}^{\a}$. 
The supercovariant connection one--form ${\omega_a}^b = dx^m 
{\omega_{ma}}^b$ is naturally introduced, via equation \eref{9} now 
evaluated
at $\theta = 0 = d \theta$, as
\beq
\label{27}
d e^a + e^{b} {\omega_b}^a = i \bpsi \G^a \psi.
\eeq
This determines $\omega$ as the metric connection, augmented by the 
standard gravitino bilinears.
The supercovariant curvature two--form is now ${R_a}^b = d 
{\omega_a}^b + {\omega_a}^c {\omega_c}^b$ with $\omega$ given in 
\eref{27}.
It is also convenient to introduce the  $A_0$, $A_2$, $A_4$
supercovariant curvatures as ($n = 1,3,5$)
\beq
F_n = R_n + C_n,
\eeq
where the $R_n$ are given in \eref{1a}-\eref{1c} and 
\eref{2a}-\eref{2c} and the 
$C_n$ are defined in \eref{7}-\eref{7b}.
More precisely
\bea
F_1 &=& \bu dV - V d \bu + C_1, \\
F_3 &=& \bu dA_2 + V d \bA_2 + C_3, \\
F_5 &=& d A_4 +i \left(A_2 d\bA_2 - \bA_2 dA_2 \right)+ C_5.
\eea

Since we write the Lagrangian as a ten--form it is also convenient to 
define the $(10-p)$--forms
\beq
E^{a_1 \ldots a_p} \equiv \frac{1}{(10-p)!} \epsilon^{a_1 \ldots 
a_p}{}_{b_1 
\ldots b_{10-p}} e^{b_1} \ldots e^{b_{10-p}}.
\eeq
In particular $E = \sqrt{-g} \, d^{10} x$.

The action for type $IIB$ supergravity with the canonical fields can 
now be written as follows:
\bea
S &=& \int E_{ab} R^{ab} + \frac{1}{3} E_{abc} \left( i \bpsi 
\G^{abc} D\psi + \hbox{c.c.}\right) + 4 E_a \left(i \bl \G^a D \l + 
\hbox{c.c.}\right) + \nonumber \\
&+& \frac{1}{4} ( F_5 * F_5 + f_4 * f_4) + \frac{1}{2} F_5 d A_4 - 
\frac{i}{2} \left(A_2 d \bA_2 - \bA_2 d A_2\right) C_5 + \frac{1}{2} 
C_5^{\l} 
C_5^{\psi} + \nonumber \\
&+& 2 \left[ \bF_3 * F_3 + \left( C_7 \bF_3 - \frac{1}{2} C_7 \bC_3 + 
c.c.\right)\right]+ \left( \frac{1}{2} \bC_7{}^{\psi} C_3^{\psi} - 2 
\bC_7^{\l} C_3^{\psi} + c.c.\right) + \nonumber \\
\label{29}
&+& 2 \left[ \bF_1 * F_1 + \left( C_9 \bF_1 - \frac{1}{2} C_9 \bC_1 + 
c.c.\right)\right] - 3 E (\bl \G^a \l) (\bl \G_a \l).
\eea

In the first line we have the kinetic terms for the metric, the 
gravitino and the field $\l$.
The second line contains the action $S_0$ of the preceding section, 
augmented by a term proportional to $C_5$ which compensates the gauge 
transformation of $F_5 dA_4$, but is $A_4$--independent, as required.
Since all the other fields are required to be invariant under the 
transformations $I)$, $II)$ and $A_4$ appears in \eref{29} only in 
the combination $S_0$, this action gives as (gauge fixed) equation of 
motion for $A_4$ just \eref{17}, i.e. \eref{13}.

The third and fourth lines in \eref{29} contain, between square 
brackets, the kinetic and interaction terms for $A_2$ and $A_0$ 
respectively.
These particular combinations are just the ones which respect the 
dualities $A_0 \leftrightarrow A_8$, $A_2 \leftrightarrow A_6$ as we 
will see in the next section.
Variation of \eref{29} with respect to $A_2$ and $A_0$ produces, as 
equations of motion, just the Bianchi identities \eref{3d}-\eref{3f} of 
section two.
The remaining terms in the action above are quartic in the fermions 
and are fixed by supersymmetry, which also fixes the relative 
coefficients of all the other terms.

The supersymmetry transformations of the fields can again be read 
from the superspace results.
Introducing the transformation parameter $\e^A = ( \e^{\a}, 
\ol{\e^{\a}}, 0 )$, the on--shell SUSY transformations of the 
component fields are given by covariantized superspace 
Lie--derivatives of the corresponding superfields, evaluated 
at $\theta = 0 = d \theta$:
\beq
\label{30}
\delta_{\e} \phi = \left[ \left( i_{\e} D + D i_{\e}\right) \phi 
\right]_{\theta = 0 = d \theta}.
\eeq
For the graviton, gravitino, $\l_{\a}$ and $(U,V)$, we get from 
\eref{9},\eref{10},\eref{11},\eref{3i} and the parametrization of 
$R_1$
\bea
\delta_{\e} e^a &=& i \left( \bpsi \G^a \e - \bar{\e} \G^a 
\psi\right), \\
\delta_{\e} \psi &=& D \e + i_{\e} D \psi, \\
\delta_{\e} \l &=& \frac{i}{2} F^a \G_{a} \bar{\e} + \frac{i}{24} 
F^{abc} \G_{abc} \e, \\
\delta_{\e} U &=& -2 V \bar{\e} \bl ,\\
\delta_{\e} V &=& -2 U \e \l.
\eea
The term $i_{\e} D \psi$ can be easily evaluated by substituting, in 
the r.h.s. of
\eref{10}, $\psi$ and $\bpsi$ respectively with $\e$ and $\bar{\e}$.

For what concerns the $p$--forms, due to gauge invariance and Lorentz 
invariance, \eref{30} would reduce simply to $\delta_{\e} A_p = i_{\e} 
dA_p$.
However, the presence of the Chern--Simons forms in 
\eref{1a}--\eref{1e} 
requires compensating SUSY transformations for the potentials $A_p$.
It is convenient to parametrize generic transformations for these 
potentials in such a way that the curvatures $S_n$ (and $R_n$) 
transform covariantly.
For later use we give here a complete list of all the combined 
transformations:
\bea
\label{32}
\delta A_2 &=& \delta_0 A_2, \\
\delta A_4 &=& \delta_0 A_4 + i \left(\delta_0 A_2 \bA_2 - \delta_0 
\bA_2 A_2\right),\\
\delta A_6 &=& \delta_0 A_6 + i A_2 \delta_0 A_4,\\
\delta A_8 &=& \delta_0 A_8 - \bA_2 \delta_0 \bA_6 + \frac{i}{2} 
\bA_2\bA_2\delta_0 A_4 - \frac{1}{4} \left( \bA_2 \delta_0 A_2 - A_2 
\delta_0 \bA_2\right) \bA_2\bA_2,\\
\label{32f}
\delta \hA_8 &=& \delta_0 \hA_8+\frac{1}{2}\left[ \bA_2 
\delta_0 A_6 - \frac{i}{2} 
A_2\bA_2\delta_0 A_4 + \frac{1}{4} \left( \bA_2 \delta_0 A_2 - A_2 
\delta_0 \bA_2\right) A_2\bA_2 - \hbox{c.c.}\right]
\eea
where $\delta_0 A_n$ parametrize generic transformations.

The corresponding (invariant) transformations for the curvatures are:
\bea
\label{33}
\delta S_3 &=& d \delta_0 A_2, \\
\delta S_5 &=& d\delta_0 A_4 + 2i \left(\bS_3 \delta_0 A_2 - S_3 
\delta_0 \bA_2\right), \\
\delta S_7 &=& d \delta_0 A_6 + i S_3 \delta_0 A_4 - i S_5 \delta_0 
A_2,\\
\delta S_9 &=& d \delta_0 A_8 - \bS_3 \delta_0 \bA_6 + \bS_7 \delta_0 
\bA_2, \\
\label{33f}
\delta \hS_9 &=& d \delta_0 \hA_8 - \frac{1}{2}\left[ \bS_3 
\delta_0 A_6 - S_7 \delta_0 \bA_2 - \hbox{c.c.}\right].
\eea
The transformations for the $R_n$ are easily obtained from their 
definitions \eref{2b}--\eref{2f}.

For supersymmetry transformations we have to choose, here for $n = 
2,4$,
\beq
\delta_0 A_n = i_{\e} S_n.
\eeq
For the curvatures, this leads  to 
\beq
\delta_{\e} S_n = (i_{\e} D + D i_{\e} ) S_n,
\eeq
and
\beq
\label{xx}
\delta_{\e} R_n = (i_{\e} D + D i_{\e} ) R_n,
\eeq
i.e., again to the covariant Lie derivative.
Expressing the $S_n$ in terms of the $R_n$, whose super--space 
parametrizations are known, in particular $i_{\e} R_n = i_{\e}( F_n - 
C_n) = - i_{\e} C_n$, the transformations \eref{xx} can be easily 
evaluated.

It remains to choose the SUSY transformation law for the auxiliary 
field $a(x)$. Since this field, being non propagating, has no 
supersymmetric partner, the simplest choice turns out to be actually 
the right 
one. We choose
\beq
\delta_{\e} a = 0.
\eeq

This concludes the determination of the on--shell SUSY transformation 
laws for the fields.
Due to the chirality condition \eref{13}, which is an equation of 
motion of the on--shell superspace approach,
some of these transformation laws could change by terms 
proportional to $F_5 - * F_5$, or equivalently, to $f_4 = i_{v} ( 
F_5 - * F_5)$.
As we will now see, SUSY invariance of the action, and therefore the 
closure of the SUSY algebra on the transformations $I)$ and $II)$, 
requires, indeed,  such modifications, but, in the present case, only for the 
gravitino supersymmetry transformation.

In practice the SUSY variation of the action \eref{29}, which is 
written as $S = \int {\cal L}_{10}$, can be performed by lifting 
formally the ten--form ${\cal L}_{10}$ to superspace, applying then 
the operator $i_{\e} D $ to ${\cal L}_{10}$ and using the superspace 
parametrizations and Bianchi identities of section two to show that 
$\delta_{\e} S = \int i_{\e}D{\cal L}_{10}$ vanishes.
The unique term for which this procedure does not work is 
$\frac{1}{4} \int f_4 * f_4$, because $a(x)$ cannot be lifted to a 
superfield; therefore the supersymmetry variation of this term has to 
be
performed "by hand".
In particular, one has to vary explicitly the vielbeins ${e_m}^a$ 
contained in $v_a$.

We give the explicit expression for the SUSY variation of the terms 
in $S$ which depend on $A_4$ and $v^a$ (the second line in 
\eref{29}).
This will be sufficient to guess the correct off--shell SUSY 
transformation law for the gravitino (the term proportional to $C_5$ 
is included to get a gauge--invariant expression)
\bea
\label{34}
&\delta_{\e}&\int \frac{1}{4} ( F_5 * F_5 + f_4 * f_4) + \frac{1}{2} 
F_5 
d A_4 - \frac{i}{2} (A_2 d \bA_2 - \bA_2 d A_2) C_5 =  \\
&=& \int i_{\e}\left[ \frac{i}{2} K_5 \frac{1}{4!} e^{a_1} \ldots 
e^{a_4} (\bpsi \G^{a_5} \psi) K_{a_5 \ldots a_1} + (C_5 - K_5) (dC_5 
+ 
2i \bR_3 R_3) - \frac{1}{2} C_5 d C_5\right]. \nonumber
\eea
In this expression $K_5$, which has only components along the bosonic 
vielbein $e^a$ , is the (self--dual) five--form given in 
\eref{26}, in particular $i_{\e} K_5 = 0$.
The peculiar feature of \eref{34} is that the five--form $F_5$ and 
the vector $v^a$ appear only in the peculiar combination
$$
K_5 \equiv F_5 + v f_4.
$$
This suggests to define the off--shell SUSY transformation for the 
gravitino by making the replacement
\beq
\label{a34}
F^+_{a_1 \ldots a_5} \to K_{a_1 \ldots a_5} 
\eeq
in \eref{10}. 
The consistency of this replacement with the closure of the SUSY 
algebra on the transformations  $I)$ and $II)$ is a consequence of the 
facts that $K_5$ is self--dual as is $F^{+}_5$, and that on--shell 
$F^+_5 = K_5$.

In conclusion, we choose for the gravitino the transformation law
\beq
\label{35}
\delta_{\e}\psi = D \e + i_{\e} (D\psi)_{F^+_5 \to K_5},
\eeq
while the transformation laws for all the other fields remain the 
ones given above.

The ultimate justification for \eref{35} stems from the fact that, 
with this choice, it can actually be checked, with a tedious and long 
but straightforward computation, that the action given in \eref{29} 
is indeed
invariant under supersymmetry.

Finally, let us notice that exactly the same replacements 
\eref{a34}, \eref{35} led to a supersymmetric action also for pure 
$N = 1$, $D = 6$ supergravity, which contains a chiral two--form 
gauge 
potential \cite{DLT}.

\section{$A_2 \leftrightarrow A_6$ duality symmetric action}

In this section we will present an action in which the 
complex six--form potential $A_6$ and its dual $A_2$ appear on the 
same footing.
This action will thus depend on $e^a$, $\psi^{\a}$, $\l_{\a}$, $A_0$,
$A_2$, $A_4$ and $A_6$.
The fields $A_2$ and $A_6$ and their curvatures are introduced as in 
\eref{1a}--\eref{1e}, \eref{2a}--\eref{2f} and \eref{3a}--\eref{3f},
 and the associated supercovariant 
curvatures in ordinary space are again given by
$$
F_n = R_n + C_n,
$$
now for $n = 0,2, 4, 6$.
Since $R_3$ and $R_7$ satisfy now their Bianchi identities 
identically, the dynamics is introduced via the duality relation 
\eref{14}
\beq
\label{36}
* F_{7} = F_3,
\eeq
which amounts now to the equation of motion for the system $(A_2, 
A_6)$; this eventually allows to eliminate $A_6$ in favour of $A_2$.
In summary, the action we search for has to give rise to the 
equation \eref{36}.

We proceed using the tools introduced in section three; we introduce 
again the scalar field $a(x)$ and the vector $v_a$ and 
define the projected forms
\bea
g_2 &\equiv& i_v ( F_3 - * F_7), \nonumber \\
g_6 &\equiv& i_v ( F_7 - * F_3).
\eea

These complex forms are $SU(1,1)$ singlets and carry $U(1)$ charge 
$+2$.
Due to \eref{24} the duality condition \eref{36} decomposes then as
\beq
\label{37a}
F_3 - * F_7 = - v g_2 + (* v g_6). 
\eeq
The projections analogous to $f_4$, instead,  are the complex forms
\bea
f_2 &\equiv& U g_2 - V \bg_2, \nonumber \\
f_6 &\equiv& U g_6 - V \bg_6,
\eea
which are $U(1)$ singlets, and $(f_2, \bff_2)$ and $(i f_6, \ol{ i 
f_6})$ are $SU(1,1)$ doublets.
The duality equation of motion \eref{36} is then equivalent to 
\beq
f_2 = 0 = f_6 \qquad \Leftrightarrow \qquad g_2 = 0 = g_6.
\eeq
It is also convenient to define
\beq
g_4 \equiv f_4,
\eeq
and,  
\beq
\label{39a}
K_3 \equiv F_3 + v g_2,
\eeq
which generalizes the analogous formula \eref{26} for $A_4$,
\beq
K_5 \equiv F_5 + v g_4.
\eeq

The action, which involves now  (apart from the fermions, the metric 
and the auxiliary field $a(x)$) the forms $A_0$, $A_2$, $A_4$, 
$A_6$, can 
be written simply as
\beq
S_{(2,6)} = S + 2 \int \bg_2 * g_2,
\eeq
where $S$ is the basic action given in the previous section.
The piece we added is invariant under global $SU(1,1)$ and local 
$U(1)$, as is $S$.

We want now to show that $S_{(2,6)}$ exhibits the following features:

{\bf i)} the fields $(A_2, A_6)$ play a symmetric role;

{\bf ii)} the dynamics associated to $S_{(2,6)}$ is equivalent to the 
dynamics described by the original action $S$; this will be shown 
through an analysis of the symmetries possessed by $S_{(2,6)}$;

{\bf iii)} $S_{(2,6)}$ is supersymmetric; to show this we have to find 
appropriate supersymmetry transformation laws for the fields.

The duality symmetry under $A_2 \leftrightarrow A_6$ is established 
by extracting from $S_{(2,6)}$ the relevant contributions depending on 
$A_2$ and $A_6$ and by rewriting them in a duality symmetric way.
These contributions are the square bracket in the third line
of \eref{29} and the added term $2 \int \bg_2 * g_2$. One finds indeed
\bea
&2& \left[ \bF_3 * F_3 + \left( C_7 \bF_3 - \frac{1}{2} C_7 \bC_3 + 
c.c. \right) + \bg_2 * g_2\right] = \\
&=& \left[ R_3 \bR_7 + \bC_7 R_3 + \bC_3 R_7 - v \left( 
\bg_6 F_3 + 
\bg_2 F_7 \right) \right] + c.c. \nonumber
\eea

A {\it completely} duality symmetric form is forbidden by the 
appearance of the Chern--Simons forms in the definition of the 
$S_n$, and hence of the $R_n$. In particular, it can be seen that the term 
$R_3 \bR_7 + \bR_3 R_7 = S_3 \bS_7 + \bS_3 S_7$, 
in the absence of Chern--Simons forms, would become a total derivative.

Now we will examine the bosonic symmetries of the action.
To this end we consider generic variations of the fields $a$, $A_2$ 
and $A_6$ and parametrize them as in \eref{32}.
The variation of the action can then be computed to be:
\bea
\delta S_{(2,6)} &=& \int -\frac{2v}{ \sqrt{-(\D a)^2}} \left( f_2 
\bff_6 + \bff_2 f_6+ \frac{1}{4} f_4 f_4 \right)d \delta a + \nonumber \\
&+& \left[ d (vf_4) - 2i v \left( \bff_2 S_3 - f_2 \bS_3 
\right)\right] 
\delta_0 A_4 + \nonumber \\
&+& \left\{ 2 \left[ d (v f_6) + i v \left( f_4  S_3 - f_2 S_5\right) 
\right] \delta_0 \bA_2 +  c.c. \right\} + \nonumber \\
\label{42}
&+& \left\{ 2 d (vf_2) \delta_0 A_6 + c.c.\right\}.
\eea

From this formula it is not difficult to realize that the action is 
invariant under the following transformations, which generalize the 
transformations $I)$ and $II)$ of section three  (n=2,4,6):
\bea
I) && \delta_0 A_n = \Lambda_{n-1} da, \qquad  \qquad \ \delta a = 
0, \\
\label{44}
II) && \delta_0 A_n = - \frac{\f}{\sqrt{- (\D a)^2}} f_n,  \qquad 
\delta a = 
\f.
\eea
From the invariance $II)$ one can conclude that $a(x)$ is again 
non propagating and its equation of motion,
\beq
d \left[ \frac{v}{\sqrt{ - (\D a)^2}} \left(f_2 \bff_6 + \bff_2 f_6 + 
\frac{1}{4} f_4 f_4\right) \right] = 0,
\eeq
can be easily  seen to be a consequence of the equations of motion 
for 
$A_2$, $A_4$, $A_6$.
These equations of motion, which can be read from \eref{42}, are, in 
fact, given by
\bea
\label{45}
d(vf_2) &=& 0, \\
\label{46}
d(vf_4) &=& 2 i v \left( \bff_2 S_3 - f_2 \bS_3\right),  \\
\label{47}
d(vf_6) &=& i v \left( f_2 S_5 - f_4 S_3\right) .
\eea
The invariances $I)$, which hold also for finite transformations, can 
be used to reduce these equations to $f_2 = f_4 = f_6 = 0 $, in the 
same way as we did in section three.
Starting from \eref{45} one can use $\l_1$ to set $f_2 = 0$.
At this point the right hand side of \eref{46} is zero and one can 
use 
$\l_3$ to set $f_4 = 0$. 
With $f_2 = 0 = f_4$, the r.h.s. of \eref{47} is also zero and 
finally one can 
use $\l_5$ to make $f_6$ vanishing.
This leads to
\bea
F_3 &=& * F_7, \nonumber\\
F_5 &=& * F_5.
\eea

The equations of motion for the other fields, with the these gauge 
fixings, are actually the same as the ones determined from $S$, the 
basic action.
This is due to the fact that the added term is {\it quadratic} in 
$g_2$, which vanishes because $f_2 = 0$.

The last issue concerns supersymmetry.
We keep for $a(x)$, $e^a$, $A_0$, $A_2$ and $A_4$ the same 
SUSY--transformation laws as in the preceding section.
In particular, $a(x)$ remains invariant and 
$$
\delta_0 A_n = i_{\e} S_n
$$
which holds now for $n=  2,4,6$ and fixes also the 
SUSY--transformation law for $A_6$, the new field.
To find the transformation laws for the fermions it is convenient to 
proceed as follows.
We extract from \eref{29} all terms which depend on $A_2$ and 
$A_4$, let us call 
them\footnote{These are just the ones in the second and third line 
of \eref{29} without the quartic terms in the fermions.} all together $\tilde{S}$.
Then we perform the SUSY variation of
\beq
\tilde{S} + 2 \int \bg_2 * g_2,
\eeq
leaving the transformations of the fermions, $\psi^{\a}$ and 
$\l_{\a}$, generic.
An explicit computation of this variation leads to the remarkable 
result that
\beq
\label{yy}
\delta_{\e}\left( \tilde{S} + 2 \int \bg_2 * g_2 \right) = \left( 
\delta_{\e} \tilde{S} \right)_{F_3 \to K_3}
\eeq
i.e. the variation of \eref{yy} coincides with the variation of 
$\tilde{S}$ if we replace in $\delta_{\e} \tilde{S}$
\beq
F_{a_1 a_2 a_3} \to K_{a_1 a_2 a_3},
\eeq
where the form $K_3$ has been defined in \eref{39a}.
The main ingredient in this computation is the identity
\beq
K_3 \equiv F_3 + v g_2 = * \left( F_7 + v g_6 \right),
\eeq
which is nothing else than \eref{37a}.
The result \eref{yy} and the fact that $S$ is invariant under 
supersymmetry imply that the action $S_{(2,6)}$ is invariant under 
supersymmetry if we choose for the fermions the transformation laws:
\bea
\delta_{\e}\psi &=& D \e + \left[ i_{\e} (D 
\psi)\right]_{\stackrel{F_5^{+} 
\to K_5}{ F_3 \to K_3}}, \\
\delta_{\e} \l &=& \left[ i_{\e} (D \l)\right]_{F_3 \to K_3} = 
\frac{i}{2} F^a \G_a \bar{\e} + \frac{i}{24} \left( K^{abc} 
\G_{abc}\right) \e.
\eea

This concludes the proof that the action $S_{(2,6)}$ provides a 
consistent, manifestly $A_2 \leftrightarrow A_6$ duality invariant 
Lagrangian formulation for $N = IIB$, $D = 10$ supergravity.
It is also manifestly invariant under the local Lorentz group 
$SO(1,9)$, under local $U(1)$ and under global $SU(1,1)$.

In the next section we extend this procedure to the dualization of 
the 
scalars $A_0 = ( U, -\bv)$.

\section{$A_0 \leftrightarrow A_8$ duality symmetric action}

The dualization of the scalars follows the strategy developed in the 
preceding section.
We will write an action in which the scalars and the eight--forms 
appear simultaneously; since the definition of the curvatures of the 
eight--forms (equations \eref{1d}--\eref{1e}) requires the presence 
of Chern--Simons forms containing $S_7 = d A_6 + \ldots$, the action 
$S_{(0,8)}$ we search for has to describe also the dynamics of $A_6$. 
Thus our starting point will be the action $S_{(2,6)}$ and 
$S_{(0,8)}$ 
will involve the whole tower of potentials and dual potentials $A_n$ 
$(n = 0,2,4,6,8)$. 
The $S_n$, $R_n$ and $F_n$ are introduced as in section two, at 
$\theta = 0 = d \theta$.

We recall that we introduce three eight--forms $(A_8, \bA_8, \hA_8)$, 
a $SU(1,1)$ triplet, and that the equations of motion, which have to 
be produced by the action $S_{(0,8)}$, are
\bea
\label{50}
F_9 &=& * F_1, \\
\label{51}
\hF_9 &=& 0.
\eea
Equation \eref{51} fixes the non propagating purely imaginary 
eight--form $\hA_8$ while equation \eref{50} transforms the Bianchi 
identities for $A_8$ in equations of motion for $A_0$ (and vice-versa).
$S_{(0,8)}$ has still to produce the duality relations $F_5 = * 
F_5$ and $F_3 = * F_7$.
We already know how to get  equation \eref{50} from a Lagrangian 
formulation, just in the same  way as we got in the preceding section 
$F_3 = * F_7$. 

On the other hand, equation  \eref{51}, which is not a duality 
relation 
between curvatures, needs a (simple) adaptation of our method: we introduce a 
new auxiliary purely imaginary one--form, which we call $\hF_1$, 
which is, however, {\it not} the differential of a scalar and 
satisfies no Bianchi identity.
Then we add to the action the term $2 \int \hF_1 * \hF_1$, which will 
eventually imply the vanishing of $\hF_1$, and impose then with our method the 
duality relations
\bea
F_9 &=& * F_1, \nonumber \\
\label{52}
\hF_9 &=&  * \hF_1.
\eea
This will finally lead to $\hF_9 = 0$. 
We choose for $\hF_1$ a vanishing $U(1)$ charge and take it to be a 
$SU(1,1)$ singlet.

The $g_0$ and $g_8$--projections of our duality relations are given by
\bea
g_0 = i_v ( F_1 - * F_9), \qquad g_8 = i_v ( F_9 - * F_1), \nonumber 
\\
\label{53}
\hg_0 = i_v ( \hF_1 - * \hF_9), \qquad \hg_8 = i_v ( \hF_9 - * \hF_1),
\eea
such that, as before
\bea
F_1 - * F_9 &=& - v g_0 + * (v g_8), \nonumber \\
\label{54}
\hF_1 - * \hF_9 &=& - v \hg_0 + * (v \hg_8).
\eea

The zero and eight--forms defined in \eref{53} are all $SU(1,1)$ 
singlets;
$g_0$ and $g_8$ have  $U(1)$ charge $+4$ and $\hg_0$ and $\hg_8$ have 
$U(1)$ charge $0$.
It is also convenient to combine these forms into forms which have all 
$U(1)$ charge zero and form $SU(1,1)$ triplets:
\bea
f_0 &\equiv& \bu^2 \bg_0 - \bv^2 g_0 - 2 \bu \bv \hg_0, \nonumber \\
\label{55a}
\hf_0 &\equiv& \bu V \bg_0 - U \bv g_0 - (|U|^2 + |V|^2) \hg_0,
\eea
and
\bea
f_8 &\equiv& \bu^2 \bg_8 - \bv^2 g_8 - 2 \bu \bv \hg_8, \nonumber \\
\label{55}
\hf_8 &\equiv& \bu V \bg_8 - U \bv g_8 - (|U|^2 + |V|^2) \hg_8.
\eea
$\hf_0$ and $\hf_8$ are purely imaginary and $f_0$ and $f_8$ are
complex. 

Defining the $SU(1,1)$ Lie--algebra elements
\bea
\label{55b}
F_0 &=& \left( \bet{cc} $\hf_0$ & $\bff_0$ \\ $f_0$ & - $\hf_0$ 
\eet\right),  \\
\label{55bb}
G_0 &=& \left( \bet{cc} -$\hg_0$ & $g_0$ \\ $\bg_0$ &  $\hg_0$ 
\eet\right),
\eea
and
\bea
F_8 &=& \left( \bet{cc} $\hf_8$ & $\bff_8$ \\ $f_8$ & - $\hf_8$ 
\eet\right),  \\
G_8 &=& \left( \bet{cc} -$\hg_8$ & $g_8$ \\ $\bg_8$ &  $\hg_8$ 
\eet\right),
\eea
the definitions  \eref{55a} and \eref{55} can also be cast in the form
\beq
F_0 = W G_0 W^{-1}, \qquad F_8 = W G_8 W^{-1}, 
\eeq
where $W$ is the scalar field matrix defined in \eref{0}. This 
makes the transformation properties of $F_0$ and $F_8$ as 
$SU(1,1)$ triplets (adjoint representation) manifest. 
The duality relations \eref{52} are then equivalent to
\beq
\label{57}
F_0 = 0 = F_8 \quad \Leftrightarrow \quad G_0 = 0 = G_8.
\eeq

The action which depends now on all the fields (and dual fields) of 
$IIB$ supergravity and also on the auxiliary fields $a$ and $\hF_1$ 
is :
\beq
S_{(0,8)} = S + 2 \int \left( \bg_2 * g_2  + \bg_0 * g_0 + \hg_0 * 
\hg_0 + \hF_1 * \hF_1 \right).
\eeq
The second and third terms in the brackets are analogous to the term 
quadratic in $g_2$, which was also present in $S_{(2,6)}$; the last 
term will eventually imply the vanishing of $\hF_1$.
Since the action is quadratic in $\hF_1$ one could think that its 
equation of motion could be substituted back in the action, leading 
to 
the elimination of the auxiliary field $\hF_1$.
This is, actually,  not the case.
In fact, under a generic variation of $\hF_1$ one has:
\beq
\label{59}
\delta S_{(0,8)} = 2 \delta \int \left(  \hg_0 * \hg_0  + \hF_1 * \hF_1 
\right) = 4 \int \left( \hF_1 + v \hg_0\right) * \delta \hF_1,
\eeq
and the equation of motion for $\hF_1$ is 
\beq
\label{60}
\hF_1 = - v i_v \left( \hF_1 - * \hF_9\right),
\eeq
which is equivalent to 
\bea
\label{61}
i_v * \hF_1 = 0 &\Leftrightarrow& v \hF_1 = 0, \\
\label{62}
i_v * \hF_9 = 0 &\Leftrightarrow& v \hF_9 = 0 .
\eea
In deriving \eref{61}--\eref{62} we used \eref{24} and the 
operator identity
\beq
i_v * = - * v
\eeq
which holds on any $p$--form.
Therefore, the component of $\hF_1$ parallel to $v$ remains 
undetermined and, moreover, one has the constraint \eref{62} on 
$\hF_9$.
For this reason the auxiliary field $\hF_1$ can not be eliminated from 
the action.

When one takes the equations of motion for the eight--forms into 
account and fixes the symmetries $I)$ and $II)$ (to be discussed now) 
then, as we will see below, one arrives actually to the equations 
\eref{57} (i.e. \eref{52}), and then \eref{60} implies indeed $\hF_1 
=0$ and $\hF_9 = 0$, which is precisely what we want.

Now we discuss the bosonic symmetries exhibited by the action 
$S_{(0,8)}$.
In this case, for the transformations of the type $I)$ it is 
convenient to proceed in a slightly different manner from the 
preceding section.
We consider generic variations $\delta A_n$ $(n = 0, \ldots, 8,\hat{8})$, 
including now also the scalars, and take for $\hF_1$ the 
transformation law
\beq
\label{64}
\delta \hF_1 = -2 i \delta Q,
\eeq
where $Q$ is the $U(1)$ connection (the fermions, the metric and the 
field $a(x)$ are not varied).
Then the variation of the action can be expressed in terms of the 
variations of the curvatures as follows:
\beq
\label{65}
\delta S_{(0,8)} = - \int v \left[ 2 \sum_{n= 0,2,6,8} \left( \bg_n 
\delta R_{9-n} + c.c. \right) + g_4 \delta R_5 + 4 \hg_0 \delta 
\hR_9 - 8 i \hg_8 \delta Q \right].
\eeq
This vanishes for the transformations ($n= 2,4,6,8,\hat{8}$)
\beq
\label{66}
Ia) \qquad \delta_0 A_n = \l_{n-1} da, \qquad \delta U = 0 = \delta V.
\eeq
The reason is that, since $\delta U = 0 = \delta V$, $\delta R_n$ is 
given by a linear combination of the $\delta S_n$, and from 
\eref{33}--\eref{33f} 
one sees that under $Ia)$ all $\delta S_n$ are proportional to $ v = 
\frac{1}{\sqrt{-(\D a)^2}} da$. 
Since in \eref{65} all $\delta R_n$ are multiplied by $v$ they drop 
out.
On the other hand, for $\delta U = 0 = \delta V$ we have $\delta Q = 
0$.

It remains to find the symmetries of the type $I)$ for the scalars.
The formula for $\delta_0 A_n$ in \eref{66} cannot be applied directly 
to the scalars, but we can observe that, due to gauge invariance, for 
example for $n = 2$, the transformation \eref{66} is equivalent to 
$\delta A_2 = d \l_1 a$, or
$$
\delta A_2 = \Omega_2 a, \quad d \Omega_2 = 0.
$$
This suggests that the transformations for the scalars analogous to 
\eref{66}, should shift them by $a(x)$ multiplied by a constant, or 
more generally, by some functions of $a(x)$.
To be more precise, since the fields $U$ and $V$ are restricted by 
the 
condition $|U|^2 - |V|^2 = 1$, we can parametrize generic 
transformations of these fields by local infinitesimal $SU(1,1)$ 
transformations
\bea
\delta U &=& \g \bv + \b U, \nonumber \\ 
\label{67}
\delta V &=& \g \bu + \b V, 
\eea
where $\g$ is complex and $\b$ purely imaginary; the matrix 
$$
M \equiv \left(\bet{cc} $\b$ & $\g$ \\ $\bar{\g}$ & $-\b$ \eet \right)
$$
is indeed an element of the Lie--algebra of $SU(1,1)$.
Now, the action $S_{(0,8)}$ is manifestly invariant under global 
$SU(1,1)$ transformations of the scalars {\it and} the forms $A_n$, 
i.e. when $\b$ and $\g$ are constants and one accompanies \eref{67} 
with the corresponding global $SU(1,1)$ transformations of $A_2$, 
$A_6$ and $(A_8, \bA_8, \hA_8)$.
This suggests to define the following transformations of the type 
$I)$ 
for the scalars (which comprise now also compensating transformations 
for the forms $A_n$) as
\bea
\delta U &=& \g \bv + \b U, \nonumber \\
\delta V &=& \g \bu + \b V, \nonumber \\
\delta A_2 &=& -\g \bA_2 + \b A_2, \nonumber \\ 
Ib) \qquad \qquad \qquad \delta A_4 &=& 0,  \nonumber \\
\delta A_6 &=& \g \bA_6 + \b A_6, \nonumber \\
\delta A_8 &=& 2\left(\g \hA_8 - \b A_8\right),\nonumber \\
\delta \hA_8 &=& \g A_8 - \bar{\g} \bA_8, \nonumber 
\eea
where $\g$ and $\b$ are now arbitrary functions of $a$, i.e. $\b = 
\b(a)$, $ \g = \g(a)$ (In this case we do not use the 
parametrizations \eref{32}--\eref{32f}).
The transformations for $A_2$ and $A_6$ and $(A_8, \bA_8, \hA_8)$ are 
just their transformations as $SU(1,1)$ doublets and triplet 
respectively, with $a$--dependent transformation parameters; the 
transformations $Ib)$ constitute therefore  a "quasi--local" 
$SU(1,1)$ transformation for all the fields where the transformation 
matrix $M$ depends on $x$ only through the function $a(x)$: $M = 
M(a)$. It is understood that for $\hF_1$ we choose the transformation
\eref{64}. 
 
From \eref{65} it can now be seen that $S_{(0,8)}$ is invariant under 
$Ib)$.
Since the $R_n$ and $Q$ are all invariant under global $SU(1,1)$ 
transformations, under the quasi--local transformations $Ib)$ we 
have that $\delta R_n$ and $\delta Q$ are all proportional to $da$, 
i.e. to $v$.
For example 
\bea
\delta Q &=& \frac{1}{2i} \left[ \bu\bv \g^\prime - UV 
\bar{\g}^\prime + \left( |U|^2 + |V|^2\right) \b^\prime \right] da, \\
\delta R_3 &=&  \left[ \left( \bu\b^\prime - V 
\bar{\g}^\prime\right) A_2 - \left( V\b^\prime + \bu 
\g^\prime\right) \bA_2 \right] da,
\eea
where $\b^\prime ={d\b\over da}, \g^\prime ={d\g\over da}$.
Therefore $\delta S_{(0,8)}$ vanishes because of the presence of 
another factor $v$ in \eref{65}.

Due to the fact that the $a$--dependent  $SU(1,1)$ transformations 
form a group we can conclude that $S_{(0,8)}$ is invariant also under 
any {\it finite} $a$--dependent $SU(1,1)$ transformation.

For what concerns the transformation $II)$, which should allow to 
eliminate the auxiliary field $a(x)$, we proceed as in the preceding 
section.
The general variation of $S_{(0,8)}$, parametrized by the 
variations $\delta_0 A_n$ according to \eref{32}, by the 
transformations of the scalars \eref{67} and by a generic variation 
of 
the field $a$ is:
\bea
\delta S_{(0,8)} &=& \int -\frac{2 v}{\sqrt{-(\D a)^2}}\left[ f_0 
\bff_8 + \bff_0 f_8 + 2 \hf_0 \hf_8 + f_2 \bff_6 + \bff_2 f_6 + 
\frac{1}{4} f_4 f_4\right] d \delta a + \nonumber \\
&+& \left[ d(vf_4) - 2i v(\bff_2 S_3 - f_2 \bS_3)\right] \delta_0 
A_4 + \nonumber \\
&+& \left\{ 2 \left[ d(vf_6) + v \left( i f_4 S_3 - i f_2 S_5 - 
\bff_0 \bS_7 - 
\hf_0 S_7 \right)\right]\delta_0 \bA_2 + c.c. \right\} + \nonumber \\
&+& \left\{ 2 \left[ d(vf_2) + v \left( \bff_0 \bS_3 - \hf_0 
S_3\right)\right] 
\delta_0 \bA_6 + c.c. \right\} + \nonumber \\
&+& 2 \left[ d(vf_8) + v \left( 2 \hf_0 S_9 - 2 f_0 \hS_9 - \bff_6 
\bS_3 + \bff_2 \bS_7 + c.c. \right) \right] \g + \nonumber \\
&+& 2 \left[ 2 d(v\hf_8) + v \left( 2 \bS_9 f_0 - \bS_3 f_6 - \bS_7 
f_2 - c.c.\right)\right] \b \nonumber \\
\label{A1}
&+&  \left[  2 d (v \bff_0) \delta A_8 + c.c.\right] + 4 d(v \hf_0) 
\delta \hA_8. 
\eea

From this formula, which generalizes \eref{42}, one sees that 
$S_{(0,8)}$ is indeed invariant 
under the transformations of the type $II)$ given by
\bea
\delta a &=& \f, \nonumber \\
\delta_0 A_n &=& - \frac{\f}{\sqrt{-(\D a)^2}} f_n, \qquad \hbox{(n = 
2,4,6,8,$\hat{8}$)}, 
\nonumber \\
II) \qquad \qquad \delta U &=& - \frac{\f}{\sqrt{-(\D a)^2}} 
(U \hf_0 + \bv \bff_0), \\
\delta V &=& - \frac{\f}{\sqrt{-(\D a)^2}} (V \hf_0 + \bu \bff_0), 
\nonumber 
\eea 
where for $\hF_1$ we choose again \eref{64} as transformation law. 
These transformations allow again to eliminate the auxiliary field 
$a(x)$.

The transformations $I)$, instead, allow to reduce the equations of 
motion for the
bosons to the first order duality equations \eref{13}--\eref{16}.
From the general variation of $S_{(0,8)}$ \eref{A1} one can, in fact, 
read the equations of motion 
for $A_8, \hA_8, A_6, A_4, A_2$ and the scalars which are respectively
given by:
\bea
\label{70}
d(v f_0) &=& 0 = d(v \hf_0), \\ 
\label{71}
d(vf_2) &=& v \left( \hf_0 S_3 - \bff_0 \bS_3 \right), \\
\label{72}
d(vf_4) &=& 2 i v \left( \bff_2 S_3 - f_2 \bS_3\right),  \\
\label{73}
d(vf_6) &=& v \left( \hf_0 S_7 + \bff_0 \bS_7 + i f_2 S_5 - i f_4 
S_3\right), \\
\label{74}
d(vf_8) &=& v \left(2 f_0 \hS_9 -2 \hf_0 S_9 +  \bff_6 \bS_3 - \bff_2 
\bS_7\right), \\
\label{75}
d(v\hf_8) &=& v \left( f_2 \bS_7 + f_6 \bS_3 - 2 f_0 \bS_9 - 
c.c.\right).
\eea

To reduce these equations to the self--duality equations, one has to 
start from \eref{70}
and use the $Ib)$ invariances in their finite form, i.e. the 
$a$--dependent $SU(1,1)$ transformations.
In more detail, \eref{70}, which can also be written as
\beq
d(v F_0) = 0,
\eeq
where $F_0$ is the $SU(1,1)$ Lie--algebra valued matrix given in 
\eref{55b}, has the general solution
\beq
\label{77}
F_0 = \sqrt{-(\D a)^2}\cdot \Sigma(a),
\eeq
where  $\Sigma(a)$ is an arbitrary Lie--algebra valued matrix which 
depends on $x^m$ only through $a(x)$. 
We want now to show that the r.h.s. of \eref{77} can be cancelled by 
a finite $a$--dependent $SU(1,1)$ 
transformation $\Lambda(a)$ (in the doublet representation of  
$SU(1,1)$ ).
To see in which way $F_0$ transforms we remember that $F_0 = W G_0 
W^{-1}$ and observe that, due to its definition \eref{55bb} and to
\eref{2aa}
\beq
\label{78}
G_0 = i_v \left[ W^{-1} dW - (\hF_1 + 2iQ) \left(\bet{cc} 1 & 0 \\ 0 
& $-1$ \eet \right)
- * \left( \bet{cc} $-\hF_9$ & $F_9$ \\ $\bF_9$ & $\hF_9$ \eet 
\right) \right].
\eeq

Thanks to \eref{64} the second term in this expression is invariant as 
is also the last one because, under
$\l(a)$, $(F_9, \bF_9, \hF_9)$ transform by terms proportional to 
$v$, and one has the operator identity $i_v* = -*v$.

Under a $SU(1,1)$ transformation we have $$ W \to \l(a) W, $$ and 
therefore,
from the first term in \eref{78} $$ G_0 \to G_0 -  \sqrt{-(\D a)^2} 
W^{-1} \left( \l^{-1} 
\frac{\delta \l}{\delta a} \right) W.$$
This leads for $F_0$ to the transformation  $$ F_0 \to \l F_0 \l^{-1} 
-  \sqrt{-(\D a)^2}  \left( \frac{\delta \l}{\delta a} \l^{-1} \right) $$
and \eref{77} transforms into 
\beq
\l F_0 \l^{-1} = \sqrt{-(\D a)^2}  \left[\Sigma(a) +  \frac{\delta 
\l(a)}{\delta a} \l^{-1}(a) \right].
\eeq
Since $ \frac{\delta \l}{\delta a} \l^{-1}$ belongs to the 
Lie--algebra of $SU(1,1)$, as
does $\Sigma (a)$, the equation
\beq
\label{79}
 \frac{\delta \l(a)}{\delta a} \l^{-1}(a) = - \Sigma (a)
\eeq
fixes, indeed, consistently and uniquely $\l (a)$, modulo a global 
$SU(1,1)$ transformation.
Choosing, therefore, $\l(a)$ as in \eref{79} we get $F_0= 0$, i.e.
\beq
\label{80}
f_0 = 0 = \hf_0.
\eeq
Since the transformation $\l(a)$ leaves the equations of motion 
\eref{71}--\eref{75}
invariant (or sends them into linear combinations), \eref{71} becomes 
now, due to \eref{80}
$$ d(vf_2) = 0.$$
This has the solution:
\beq
\label{81}
vf_2 = da d\tilde{\l}_1,
\eeq
and the transformations $Ia)$, with only $\l_1$ non--vanishing and 
$\l_1 = \tilde{\l}_1$ reduce
\eref{81} to $f_2 = 0$. 
This transformation leaves \eref{72}--\eref{75} invariant and 
maintains also \eref{80}.
Now one can proceed with \eref{72}, whose r.h.s. is now vanishing, to 
set also $f_4$ to zero and so on.
In this way one can use the transformations $Ia)$ and the equations 
of motion
\eref{70}--\eref{75} to have finally for {\it all the values of} $n$ $f_n = 0$, or, 
equivalently
\beq
g_n = 0.
\eeq
But this is equivalent to the duality relations
\bea
n=4 && F_5 = * F_5, \nonumber \\
n=2,6 && F_7 = * F_3,  \nonumber\\ 
n=0,8 && F_9 = * F_1, \nonumber \\
n=\hat{0},\hat{8} &&  \hF_9 = * \hF_1. \nonumber
\eea
The equation of motion for $\hF_1$ given in \eref{60}, i.e. $\hF_1 = 
-v \hg_0$,
leads then finally to 
$$
\hF_9 =0 =  \hF_1.
$$

This concludes the proof that the dynamics described by the action 
$S_{(0,8)}$ is equivalent to the dynamics of 
$D = 10$, $IIB$ supergravity.

The last issue is supersymmetry.
The supersymmetry variation of  $S_{(0,8)}$ can again be performed in 
a standard way, 
following the procedure of section five.
The SUSY--variation of $e^a$ is the standard one and again we choose 
$\delta_{\e} a = 0$.
The variation of the forms $A_n$ $(n = 2,4,6,8,\hat{8})$ is 
parametrized by \eref{32} with
$$ \delta_0 A_n = i_{\e} S_n, $$
as in the preceding section.
The supersymmetry transformations of the fermions are obtained from 
the standard ones, given
in section four, through the replacements:
$$
F_3 \to K_3 \equiv F_3 + v g_2, 
$$
$$
F_1 \to K_1 \equiv F_1 + v g_0. 
$$
More precisely:
\bea
\delta \psi &=& D\e + \left( i_{\e} D \psi\right)_{\stackrel{F_5^{+} 
\to K_5}{ F_3 \to K_3}}, \\
\delta \l &=& \left( i_{\e} D \l\right)_{\stackrel{F_3 
\to K_3}{ F_1 \to K_1}} = \frac{i}{2} (K^a \G_a) \bar{\e} + 
\frac{i}{24} \left(K^{abc} \G_{abc}\right) \e.
\eea 
Notice that the modification of the supersymmetry transformation laws 
are proportional to the 
$g_n$, and vanish therefore on--shell, and that $\hg_0$ does not 
enter in these modifications 
since its partner $\hF_1$ is not present in the standard formulation 
of $IIB$ supergravity.

It remains to fix the SUSY transformation of $\hF_1$.
To this order we remember that under a generic variation of $\hF_1$ 
we have, from \eref{59}
\beq
\delta S_{(0,8)} = 4 \int \hat{K}_1 * \delta \hF_1,
\eeq
where
\bea
\hat{K}_1 &\equiv&  \hat{F}_1 + v \hg_0, \\
\delta  \hat{F}_1 &=& dx^{m} \delta  \hat{F}_m.
\eea
This means that $\delta_{\e}  \hat{F}_1$ can be used to cancel from 
$\delta_{\e} S_{(0,8)}$
all "residual" terms which are proportional to $ \hat{K}_1$, i.e. terms 
which are proportional
to the equations of motion of $ \hat{F}_1$.
Such residual terms are, actually, present and it turns out that one 
has to choose
\beq
\label{84}
\delta_{\e}  \hat{F}_1 = \left[ - W_{(ab)} + \frac{1}{2} \eta_{ab} 
W_c{}^c \right] e^a \hat{K}^b
- 2v \left[ \bg_0 (\e\l) - g_0 (\bar{\e}\bl)\right],
\eeq
where $W_{ab} = i \left[ \bpsi_a \G_b \e - \bar{\e} \G_b \psi_a \right]$.
With this SUSY transformation for $\hat{F}_1$ and the transformations 
for the other fields defined above
one can check that $S_{(0,8)}$ is indeed supersymmetric:
$$ \delta_{\e} S_{(0,8)} = 0.$$

The on--shell consistency of \eref{84} can be verified as 
follows.
The equation of motion for $ \hat{F}_1$ is 
$$\hat{K}_1 \equiv \hat{F}_1 + v \hg_0 = 0$$ 
and \eref{84} reduces to:
\beq
\label{85}
\delta  \hat{F}_1 = -2  \frac{da}{\sqrt{ - (\D a)^2}} \left( \bg_0 
(\e\l) - g_0 (\be\bl)\right).
\eeq
On the other hand $\hat{K}_1 = 0$ gives
\beq
\label{86}
\hF_1 = -  \frac{da}{\sqrt{ - (\D a)^2}} \hg_0.
\eeq
Since on--shell we have
$$
G_0 = \left(\bet{cc} $-\hg_0$ & $g_0$ \\ $\bg_0$ & $ \hg_0$ \eet 
\right) = 
\sqrt{ - (\D a)^2}\cdot W^{-1} \Sigma(a) W
$$
the factor $\sqrt{ - (\D a)^2}$ disappears in \eref{85} and \eref{86}.
Since, moreover, $\delta_{\e} a = 0$, in computing the variation of 
the r.h.s.
of \eref{86} one has only to vary the fields $U$ and $V$ contained in 
$W$, and the
result can easily be seen to coincide with \eref{85}.

\section{Concluding remarks}

In this paper we presented an action which produces correctly the 
dynamics of $D=10, N=IIB$ supergravity, \eref{29}, and possesses all
relevant symmetries. A non manifestly Lorentz invariant action can be 
obtained by setting, for example, $a(x)=n_mx^m$; this eliminates the unique
auxiliary field but leaves the action still invariant under the 
transformations of the type I). These transformations allow to recover
the self--duality condition for the chiral four--form. Apart from 
fundamental reasons, the
knowledge of this action could for example be useful when one couples
the relevant $Dp$--branes to $IIB$ supergravity, i.e. considers an
action of the form $S_{IIB}^{Sugra}$ + $S_{Dp}^{\sigma-model}$.

We presented also actions which are manifestly invariant under the
interchange of the basic forms and their Hodge--duals. 
In this case the dual fields can be eliminated upon gauge--fixing 
the transformations I) while it is not possible to eliminate the basic 
fields in favour of the dual ones. With the gauge fixings we performed in
the paper, one is essentially forced to interpret the Bianchi identities
of the dual fields as equations of motion of the basic fields. As we will
sketch now, however, in a perturbative treatment one can actually
treat the basic fields and their duals in a symmetric fashion.
We will outline the procedure in the case of $A_2\leftrightarrow A_6$, 
at the linearized level, i.e. for vanishing fermions, for a flat metric
and choosing for simplicity $U=1$ and $V=0$.

First of all we choose a gauge fixing for the transformations II) according
to $a=n_mx^m$. This implies 
\beq
v_m=n_m, \qquad dv=0.
\eeq
The equations of motion for $A_2$ and $A_6$, eqs. \eref{45} and \eref{47},
reduce then to 
\bea
\label{eq1}
v \, di_v(dA_2-*dA_6)&=&0,\\
\label{eq2}
v \, di_v(dA_6-*dA_2)&=&0.
\eea
Appropriate gauge fixings for the transformations I) and for the 
gauge invariances $\delta A_n=d\Lambda_{n-1}$ are
\bea
\label{gf1}
d*A_2=&0&=d*A_6,\\
\label{gf2}
i_vA_2=&0&=i_vA_6.
\eea
This reduces \eref{eq1} and \eref{eq2} further to
\beq
\label{red}
T\D_v A_2+(\quadratello+\D_v^2)A_6=0=T\D_v A_6+(\quadratello+\D_v^2)A_2,
\eeq
where $\D_v=v^m\D_m$ and $T=*vd$. In particular, the operator $T$ satisfies,
on forms which are constrained by \eref{gf1} and \eref{gf2},
$$
T^2=\quadratello+\D_v^2.
$$
Using this one can combine the equations in \eref{red} to
get 
$$
\quadratello A_2=0= \quadratello A_6,
$$
which are the correct equations for massless fields,
and
\beq
\label{pol}
TA_2+\D_v A_6=0=TA_6+\D_v A_2,
\eeq
which is a residual constraint on the polarizations. To see which 
polarizations survive we go to momentum space, $A_n(x)\rightarrow a_n(p)$,
and choose $n^m=(1,0,\cdots,0)$. Splitting the ten--dimensional index $m$ as
$m=(0,r)$, \eref{gf2} says that only space--like indices survive in the 
polarizations and \eref{gf1} gives the transversality conditions 
\beq
\label{trans}
p^ra_{rs}=0=p^ra_{rr_1r_2r_3r_4r_5}.
\eeq
Therefore, both $a_2$ and $a_6$ carry 28 degrees of freedom, but the residual 
conditions \eref{pol} identify now these polarizations giving in momentum
space the single constraint
$$
a_{r_1\cdots r_6}={1\over 2}\varepsilon_{r_1\cdots r_6}{}^{s_1s_2s_3}
{p_{s_1}\over |\vec p|}  a_{s_2s_3}.
$$

This constitutes just a check of the fact that our $A_2\leftrightarrow A_6$
duality invariant action propagates the correct degrees of freedom. 

On the 
other hand, the gauge fixings \eref{gf1},\eref{gf2} are symmetric under
$A_2\leftrightarrow A_6$ and they could also be used in a 
functional integral. Clearly, for ten dimensional theories a functional
integral is of very little relevance; let us mention, however, that using
these gauge--fixings one can indeed perform the (duality symmetric)
integration over the two gauge fields in the four--dimensional
duality symmetric action for Maxwell's equations coupled to electric
and magnetic sources proposed in \cite{BerMed}, and the resulting
effective interaction for electric and magnetic charges turns out to be the 
expected one and, in particular, independent on $v$ \cite{KL}.

Just the same gauge--fixings
\eref{gf1},\eref{gf2} appear also appropriate in ten dimensions
for what concerns the determination
of the Lorentz--anomaly due to $A_4$, in a perturbative functional 
integral approach.

\bigskip

\paragraph{Acknowledgements.}
\ We are grateful to P. Pasti and D. Sorokin for interest 
in this work and useful discussions. This work was supported by the 
European Commission TMR programme ERBFMPX-CT96-0045 to which the 
authors are associated. 


\end{document}